# Electrical detection of the flat band dispersion in van der Waals field-effect structures

Gabriele Pasquale[1,2*], Edoardo Lopriore[1,2*], Zhe Sun[1,2], Kristiāns Čerņevičs[3], Fedele Tagarelli[1,2], Kenji Watanabe[4], Takashi Taniguchi[5], Oleg V. Yazyev[3], Andras Kis[1,2§]

[1]*Institute of Electrical and Microengineering, École Polytechnique Fédérale de Lausanne (EPFL), CH-1015 Lausanne, Switzerland*
[2]*Institute of Materials Science and Engineering, École Polytechnique Fédérale de Lausanne (EPFL), CH-1015 Lausanne, Switzerland*
[3]*Institute of Physics, École Polytechnique Fédérale de Lausanne (EPFL), CH-1015 Lausanne, Switzerland*
[4]*Research Center for Functional Materials, National Institute for Materials Science, 1-1 Namiki, Tsukuba 305-0044, Japan*
[5]*International Center for Materials Nanoarchitectonics, National Institute for Materials Science, 1-1 Namiki, Tsukuba 305-0044, Japan*

*\* These authors contributed equally to this work.*
*§ Correspondence should be addressed to: Andras Kis (andras.kis@epfl.ch)*

**Two-dimensional flat-band systems have recently attracted considerable interest due to the rich physics unveiled by emergent phenomena and correlated electronic states at van Hove singularities. However, the difficulties in electrically detecting the flat band position in field-effect structures are slowing down the investigation of their properties. In this work, we employ Indium Selenide (InSe) as a flat-band system due to a van Hove singularity at the valence band edge in a few-layer form of the material without the requirement of a twist angle. We investigate tunneling photocurrents in gated few-layer InSe structures and relate them to ambipolar transport and photoluminescence measurements. We observe an appearance of a sharp change in tunneling mechanisms due to the presence of the van Hove singularity at the flat band. We further corroborate our findings by studying tunneling currents as a reliable probe for the flat-band position**

**up to room temperature. Our results create an alternative approach to studying flat-band systems in heterostructures of 2D materials.**

## INTRODUCTION

Flat-band systems have recently become an exciting playground for studying strongly correlated phenomena and emergent physics in two-dimensional materials. After the discovery of correlated insulating and superconducting states in twisted bilayer graphene[1], the quest for materials systems exhibiting exotic phenomena has quickly spread to twisted bilayer TMDCs, where the moiré potential can induce topological states in the heterostructure[2,3]. Another material that has a flat band in its pristine form is indium selenide (InSe)[4], a layered 2D semiconductor belonging to the III-VI metal mono-chalcogenide family. It possesses distinctive electrical[5,6] and optical properties[7,8], making it an attractive platform for optoelectronic devices based on 2D materials[9–12].

The bandgap of γ-InSe changes from a direct band gap of ~1.27eV for bulk to a ~2.8eV indirect band gap in the monolayer limit[5,13,14]. The indirect-to-direct bandgap transition is due to a shift of the valence band maximum (VBM) from the Γ point towards the K points. This band structure modification is accompanied by an appearance of a flat-band dispersion and a van Hove singularity in the density of states, and has been predicted to possibly induce ferromagnetism or superconductivity[15–17]. The possibility to study emergent phenomena in devices based on InSe is appealing because of a simpler device fabrication, which does not require twist-angle engineering, unlike in other platforms based on 2D materials[1–3]. While several reports have confirmed flat-band dispersions in 2D systems, only a handful of technically demanding methods such as ARPES and STM[4,18] have so far been employed to experimentally verify the presence of a singularity in the density of states. In particular, in the case of InSe, the flat valence band has not yet been detected using electrical transport measurements, and its presence has only been revealed by STS[19] and ARPES[4]. The difficulty of detecting the flat band through electrical transport measurements is a hurdle for studying emergent physics in 2D materials.

Here, we use measurements of out-of-plane tunneling currents in hBN-encapsulated few-layer InSe samples to detect the Fermi level reaching the valence band edge via electrical measurements. The van Hove singularity provides a sudden availability of carriers that induces an abrupt change in the tunneling processes across the hBN insulating barrier. We provide evidence for the electrical detection of the flat-band position in InSe by comparing tunneling currents with lateral transport and photoluminescence (PL) measurements. The significant asymmetry of the electron and hole effective masses in few-layer InSe (e.g. $0.131m_0$ and $0.913m_0$ for 6 layers), together with the large Schottky barrier for holes (Supplementary Note 1), is found to be responsible for the lack of p-type current in few-layer InSe transport measurements reported to date[5]. In our devices, we achieve p-type conduction for different thicknesses of InSe (3 and 6 layers) and show that the onset of the out-of-plane tunneling current is consistent with the beginning of hole transport at cryogenic temperatures. We observe that lateral hole transport deteriorates for a decreasing number of InSe layers due to the high hole effective mass at the valence band. This confirms lateral transport to be an ineffective way to probe flat-band physics when going toward few-layer systems with large valence band dispersions. Instead, the magnitude of out-of-plane currents depends on the hole effective mass in the dielectric medium and not the semiconductor, while being directly related to the divergent DOS. This allows us to decouple the high availability of states from the large effective mass that hinders lateral transport, making tunneling current measurements a practical approach for detecting the van Hove singularity. Furthermore, we directly relate different regimes in the gate-dependent tunneling photocurrents to the behavior of excitonic species in the PL spectra. In particular, we focus on the bandgap renormalization effects appearing in the presence of a Fermi sea when the Fermi level is located inside the valence band[20]. The chosen range of InSe thicknesses (3 to 6 layers) allows us to directly relate the tunneling mechanisms to the lateral hole transport and PL in the flat band regime[4].

Based on our experimental evidence and theoretical models, we outline a complete picture that allows us to investigate a flat-band system and the accompanying evolution of the density of states as a function of thickness. Our work opens up the exploration of 2D flat-band systems and their emergent properties by employing tunneling photocurrents through dielectric barriers.

## RESULTS

**Device structure and band diagrams of few-layer InSe**

To obtain high-quality samples, we encapsulate γ-phase InSe flakes in hBN and use few-layer graphite (FLG) flakes as electrodes. We use a FLG bottom gate to modulate the carrier density in the semiconductor. The complete device schematic is reported in Figure 1b, representing an optically excited heterostructure based on InSe. In Figure 1a, an example of a 3-layer InSe device (3L) is shown in false color to highlight the stacked layers. The band structure for 3 layers of InSe was obtained by DFT calculations accounting for the presence of selenium vacancies ($V_{Se}$), the most common type of defects in few-layer InSe due to their low formation energy[21,22] (Supplementary Note 1). These defects produce two peaks in the density of states (DOS) close to the conduction band (CB) and a peak at ~150meV from the valence band (VB), as shown in Figure 1c and Figure 1d. In few-layer InSe, a sharp DOS increase characterizes the valence band maximum position (Figure 1e) due to the appearance of a van Hove singularity.

As outlined in the following sections, the presence of defect states dictates the tunneling behavior while the Fermi level is kept within the bandgap, in agreement with the measured PL features. By taking defect states into account, we explain the nature of lateral hole transport in InSe devices and the change in tunneling behavior at the point corresponding to the flat-band position. Moreover, our calculations reveal valence band offsets (VBOs) on the order of hundreds of meVs in heterostructures with FLG-contacted few-layer InSe on top of hBN (Figure 1c, Supplementary Figure S1), unveiling a significant asymmetrical alignment when compared to the band offsets at the conduction band (~3.8 eV). This asymmetry is crucial for the electrical detection of the DOS at the van Hove singularity at the VBM.

**Lateral transport and tunneling photocurrent**

On Figure 2a we present the gate dependence of the lateral current in our 3 and 6-layer devices (3L and 6L). We observe ambipolar transport with dominant n-type transport[5,23] while p-type conduction is strongly suppressed due to relatively high hole effective masses compared to the electron ones (Supplementary Table 1). A visible decrease in the p-type subthreshold slope is seen in the 3L with respect to the 6L device, consistent with the reduction of the hole effective mass with an increasing number of InSe layers, as discussed in Supplementary Notes 1 and 2.

Illumination of the device with a laser beam (λ = 633 nm) results in the appearance of an out-of-plane tunneling current between the InSe layers and the FLG gate, shown on Figure 2b for the 6L device, together with the lateral current as a function of gate voltage. The tunneling current shows a rapid increase at $V_G = -3.5V$, closely matching the hole conduction threshold voltage $V_{TH,p} = -3.6V$. Photocurrents for devices with varying thicknesses (3, 5, and 6 layers) are shown in Figure 2c on a logarithmic scale, revealing two main regions with linear and logarithmic trends, respectively. This behavior is reminiscent of the separation between direct tunneling (DT) and Fowler-Nordheim tunneling (FNT) mechanisms in electronic devices such as metal-insulator-metal (MIM) diodes[24,25] and can be modeled using Simmons' approximation[24,26] (Supplementary Note 3). The FNT part allows us to extract the tunnel barrier heights. We refer to the position of changing trends as the transition voltage $V_{TRAN}$, which for the 6L device occurs at $-3.5V$ while for the 3L sample, $V_{TRAN}^{3L}= -5.8V$ , close to the onset of p-type conductivity $V_{TH}^{3L} = -6.3V$.

To rule out the possibility that the drastic change in the tunneling transition is due to change in the transmission probability, we consider the differential tunneling conductance $d\,I_G/d\,V_G$ normalized over the total conductance $I_G/V_G$ (Figure 2d). As previously reported[27], the appearance of a peak in the $(d\,I_G/d\,V_G)/(I_G/V_G)$ curve indicates that the rise of the tunneling signal is due to a sudden change in the material DOS (Supplementary Note 3).

**Relationship between tunneling mechanisms and excitonic features**

In order to develop an understanding of the physical origin of the sharp transition in the tunneling photocurrent in the presence of a van Hove singularity at the VBM, we compare tunneling photocurrent measurements with gate-dependent photoluminescence spectroscopy (Figure 3a), allowing us to identify different excitonic species and defect-bound states[8]. We focus our attention on the redshift of the main excitonic species in the p-type region $X_+$, together with its strong suppression of emission for $V_G^{6L} \leftarrow 3.5V$. This corresponds to the Fermi level crossing into the valence band and the accompanying bandgap renormalization (BGR) at the flat band. This voltage also coincides with the onset of the tunneling photocurrent in the 6L device (Figure 3b) and hole accumulation in the semiconductor. Similar behavior is shown in Supplementary Note 5 for the 5L device.

In Figure 3c, we show the excitation power ($P$) dependence of the tunneling photocurrent ($I$) in the p-type region ($V_G = -7V$) and within the bandgap ($V_G = -3V$), displaying a linear and a superlinear behavior, respectively. $I \propto P^\alpha$ dependence with $\alpha \sim 2$ is commonly attributed to second-order mechanisms such as exciton-exciton annihilation (EEA)[28]. For $1 < \alpha < 2$, first-order recombination pathways also play a role, mainly represented by Auger processes[29]. Here, non-radiative exciton recombination results in energy transfer to charge carriers, which become sufficiently excited to tunnel through the hBN barrier in the presence of a vertical electric field. The effective tunneling barrier extracted for the Fowler-Nordheim region lies in the range of several hundred meV for 3L, 5L, and 6L devices, with values that are comparable with the calculated valence-band offsets for the respective structures (Supplementary Figure S1).

Previous reports have shown the excitation of charge carriers through Auger recombination and vertical photocurrent due to hot carrier tunneling for other 2D materials[25,29,30]. However, no DOS-dependent sharp transition between DT and FNT regions can be identified in either our TMDC-based samples (Supplementary Note 6) or in the literature, where the increase in photocurrent is attributed to the bending of the hBN bands by an increasing electric field. A different mechanism is therefore needed in order to explain the sharp transition between DT and FNT regions in InSe. In our case, the aforementioned suppression of PL from $X_+$ , together with the reduction of the power coefficient $\alpha$ with the onset of hole conduction, both indicate that exciton densities are reduced when the Fermi level enters the valence band. Since the generation factor is fixed only by the input optical power (Supplementary Note 4), we infer that, in the p-type region, Auger recombination plays a minor role in the tunneling pathways from InSe through the hBN barrier.

Even if hot-carrier excitonic-assisted tunneling were allowed in our system, it would not be a likely origin of the sharp onset in photocurrent since the effective barrier seen by holes is comparable to the one predicted by DFT (Figure 3f). Directly photo-excited carriers have also been previously reported as a contribution to out-of-plane currents[28,31]. However, in our case, the magnitude of the extracted effective barrier indicates that tunneling carriers are probed in the vicinity of the VBM and not from energy levels deep within the valence band (Figure 3d). Based on our understanding of the power-dependent photocurrent and the BGR, we attribute the sharp onset of the photo-assisted process to a change of the main tunneling species from hot carriers assisted by recombining defect-bound excitons ($V_{TRAN} < V_G < 0V$) to directly photo-excited holes located at the VBM ($V_G < V_{TRAN}$).

**Room-temperature tunneling current**

The low exciton binding energy in InSe (~10 meV)[12], together with the desire to investigate emergent physics at the flat band[1,2], motivated initial measurements performed at low temperatures. Extending the temperature range can further enhance our understanding of the tunneling mechanisms in few-layer InSe. In Figure 4a we show the tunneling current measured without laser illumination at 80 mK (black) and 295 K (red). At 80 mK, the dark current exhibits a linear relationship with respect to $V_G$, with no observable change in tunneling mechanisms in the gate voltage range of interest. However, at room temperature we observe a sharp transition in tunneling mechanisms from DT to FNT. In fact, higher temperatures are related to an increase in tunneling probabilities and in photocurrent signal magnitudes[32]. Moreover, the room-temperature transition voltage is consistent with that obtained at low temperatures under laser illumination ($V_{TRAN}^{RT} \simeq 5V$), as highlighted in the inset of Figure 4a. This indicates that hole tunneling from the VBM is dominating both at room temperature in the dark, where thermally enhanced out-of-plane transport of holes at the van Hove singularity is directly detectable, and at low temperatures with light, where directly photoexcited carriers near the VBM are extracted thanks to high electric fields. The temperature change only affects the magnitude of the detectable signal without influencing the origin of the tunneling current. This confirms that the change in tunneling mechanisms of out-of-plane currents is due to the Fermi level reaching the van Hove singularity in the DOS at the InSe VBM, and rules out excitonic effects.

We further observe a change in the lateral transport of holes at low and high temperatures (Figure 4b). In fact, the presence of in-gap defect states influences the transport characteristics, as discussed in Supplementary Note 2. Due to the high hole effective mass (Supplementary

Note 1), it is difficult to identify the valence band edge from the onset of hole conduction in thinner InSe samples ($N \leq 3L$). This explains the higher discrepancy in the quantities of interest shown in Figure 3e between 3L and thicker samples. We can conclude that, while the onset of p-type conduction in our InSe devices is affected by the presence of defect states at room temperature, the tunneling current transition voltage remains unaffected. Such a result supports the use of tunneling currents as a reliable mechanism to electrically detect van Hove singularities of flat band systems in field-effect devices.

**DISCUSSION**

The precise detection of the flat band in field-effect structures at low temperatures is critical for exploring emergent phenomena such as magnetism or superconductivity[34,35] and correlated electron states such as Wigner crystals and Mott insulators[2]. In this work, we have studied tunneling currents in few-layer InSe and their relation to the van Hove singularity at the valence band edge of the semiconductor (Supplementary Note 3). Although tunneling photocurrents in van der Waals heterostructures have been previously observed, their exploration has been limited to photodetector applications and to studies of excitonic features in TMDCs[28–30].

Since the presence of a flat dispersion gives rise to high effective carrier masses, the identification of lateral field-effect hole transport in InSe has been difficult, hindering the identification of the valence band edge position in devices with a low number of layers. On the other hand, the magnitude of tunneling currents is related to the effective mass of the holes in the insulating layer (hBN, $0.5\ m_0$)[26], which is significantly smaller than that in InSe ($0.931\ m_0$ for 6L). High signal-to-noise ratios in tunneling photocurrents can be successfully achieved even in thin InSe (3L in Figure 2c) by simply increasing the illumination intensity, resulting in increased carrier tunneling. For example, while we see typical lateral currents in the 10 pA range (0.4 V in a 6 L device at 4.5 K), we can observe tunneling photocurrents of ~100 nA under similar conditions (Figure 3c). While no observable damage to the structure is present with tunneling photocurrents, the increase of lateral transport in the p-type region by gating is limited by dielectric breakdown[33]. Moreover, although the p-type onset changes with temperature due to defect states within the bandgap, the onset of the tunneling current remains unchanged, providing a reliable marker for the flat band position.

We have shown that the change in tunneling mechanisms is related to the excitonic properties of the semiconductor, motivating combined exploration of out-of-plane photo-assisted tunneling with spectroscopy techniques in materials possessing a van Hove singularity. One significant advantage of exploiting photo-tunneling in flat-band materials is that only two electrodes are needed to identify the position of the van Hove singularity at low temperatures, in our case the InSe contact and the bottom gate.

In light of our findings, few-layer InSe represents a reliable and highly reproducible platform to study flat-band physics in field-effect structures based on 2D materials[36]. Our results are expected to motivate the further exploration of flat-band materials and their properties by out-of-plane tunneling currents, enriching the investigation of strongly correlated phenomena, correlated insulating states, and emergent physics in van der Waals heterostructures.

**ACKNOWLEDGEMENTS**

We acknowledge helpful discussions with Juan Francisco Gonzalez Marin. We acknowledge the support in microfabrication and e-beam lithography from EPFL Centre of MicroNanotechnology (CMI) and thank Z. Benes (CMI) for help with electron-beam lithography. This work was financially supported by the Swiss National Science Foundation (grant nos. 172543, 175822, 177007 and 164015), the European Union's Horizon 2020 research and innovation programme under grant agreements 785219 and 881603 (Graphene

Flagship Core 2 and Core 3), and the Marie Curie Sklodowska ITN network "2-Exciting" (grant no. 956813). K.W. and T.T. acknowledge support from the Elemental Strategy Initiative conducted by the MEXT, Japan (Grant Number JPMXP0112101001) and JSPS KAKENHI (Grant Numbers JP19H05790 and JP20H00354). Computations were performed at the Swiss National Supercomputing Centre (CSCS) under project No. s1146 and the facilities of Scientific IT and Application Support Center of EPFL

## METHODS

### Device fabrication

The heterostructures used throughout this study are fabricated by a three-step dry transfer technique. First, the hBN and few-layer graphene (NGS) building blocks are exfoliated on silicon oxide. The bottom hBN flake is picked up with a polycarbonate (PC) membrane on PDMS and released on top of the few-layer graphene (FLG) bottom gate. The few-layer InSe (HQ Graphene) flakes are exfoliated on PDMS (Gelpak), identified by optical contrast, and transferred on top of the previously fabricated hBN/FLG stack. Top hBN and FLG contacts are subsequently picked up, carefully aligned, and released on top of the InSe flake, fully encapsulating the heterostructure. All these steps are performed inside an argon-filled glovebox (Inert) to avoid material degradation and contamination. Once capped, the sample is annealed at 340 ºC in a high vacuum at $10^{-6}$ mbar for 6 h. Finally, electrical contacts are fabricated by e-beam lithography and metal evaporation (2/100 nm Ti/Au).

### Optical and electrical measurements

All measurements shown in this work were carried out under vacuum at 4.5K unless specified otherwise. PL measurements were performed by focusing a laser on a spot of about 1 µm diameter on the sample. A narrow-linewidth tunable continuous wave laser (MSquared) is used both for PL measurements and for tunneling photocurrent measurements. The incident power was varied from 1 µW to 5 mW for power dependence measurements (Figure 3c) and kept at 50 µW for the PL spectroscopy shown in Figure 3a and tunneling photocurrent measurements unless otherwise specified. Transport measurements were carried out at room temperature, helium temperature (4.2K), and 80 mK with a Keithley 2636 sourcemeter. The 80 mK temperature was achieved inside a dilution fridge from Oxford Instruments, with a custom-made window and mirrors that allow us to perform optical and optoelectronic measurements.

**First-principles calculations**

Our first-principles calculations were performed using VASP [37] at the density-functional theory level. For all structure relaxations, semilocal PBE functional[38] was used. The conjugate gradient method was used to optimize atomic positions and lattice constants, where the total energy and atomic forces were minimized. The convergence was reached when the maximum force acting on each atom was less than 0.01 eV/Å, while the next energy step was smaller than $10^{-5}$ eV. For accurate bandgap estimations of pristine systems, the hybrid HSE06[39] was employed, while for defect calculations with large supercells, we used a modified Becke-Johnson exchange potential in combination with LDA-correlation[40,41]. Kohn-Sham wave functions were expanded in a plane-wave basis set with a kinetic energy cutoff of 400 eV, while electron-core interactions were described through the projector augmented wave (PAW) method[42,43]. All structures were subjected to periodic boundary conditions with a vacuum layer of 10 Å perpendicular to the layers to prevent interaction between replica images. The reported effective masses are obtained as an average between the different directions in the Brillouin zone, from $\Gamma$ to K, and from $\Gamma$ to M. We employed a parabolic fitting, with a sampling of four points in the k-space to obtain the result (for further information see Supplementary Note 1).

Defect calculations were carried out by building 4×4 supercells. Finally, 10×10×1 and 3×3×1 k-point mesh was used for the integration over the Brillouin zone for pristine and defective systems respectively.

## AUTHOR CONTRIBUTIONS

A.K. initiated and supervised the project. G.P. fabricated the devices. K.W. and T.T. grew the h-BN crystals. G.P., Z.S, and E.L. performed the optical and transport measurements assisted by F.T.. G.P. and E.L. analysed the data with input from Z.S. and A.K.. K.C. and O.Y. performed DFT calculations for the band structures, defect states, and band alignments. G.P. and E.L. wrote the manuscript, with inputs from all authors.

## COMPETING FINANCIAL INTERESTS

The authors declare no competing financial interests.

## DATA AVAILABILITY

The data that support the findings of this study are available on Zenodo at doi: XXXX.

# FIGURES

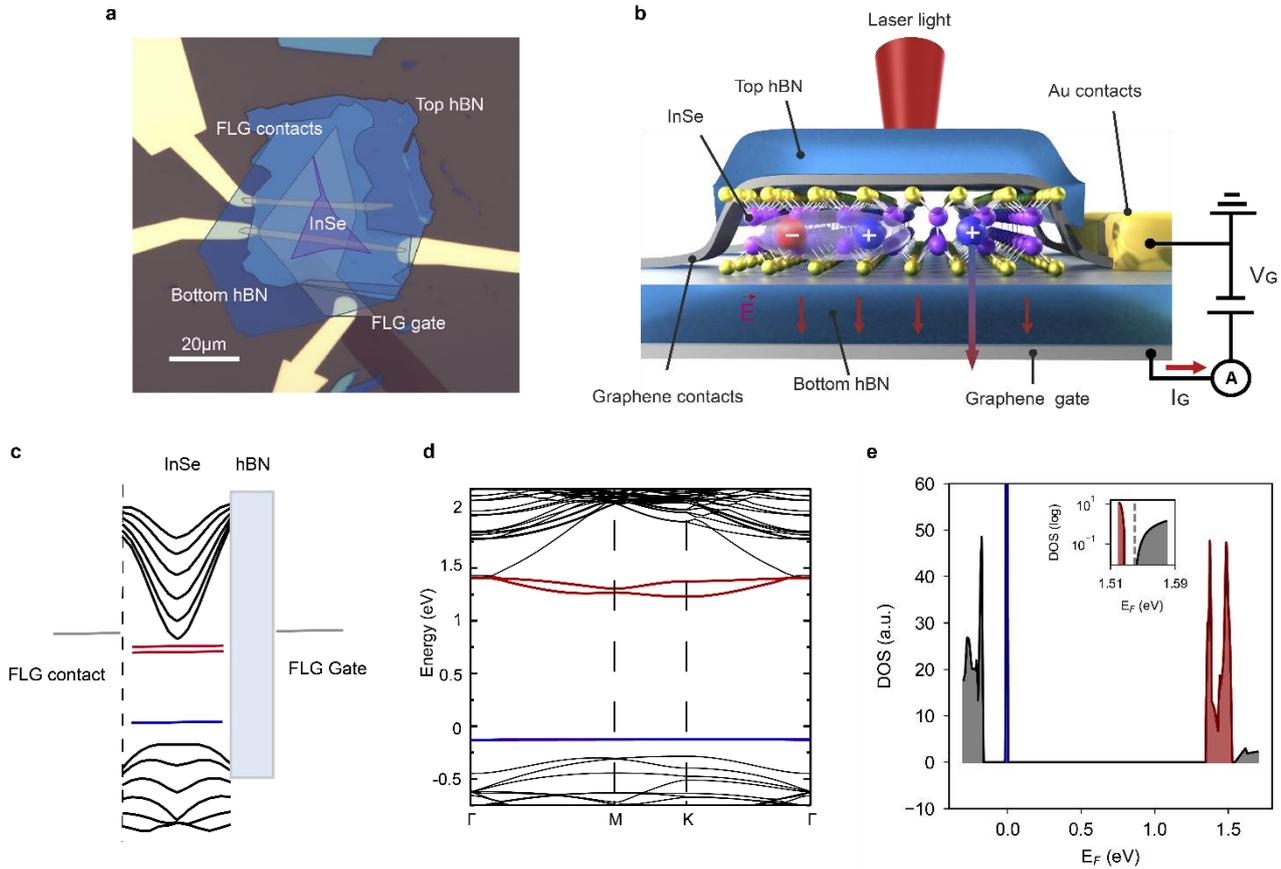

**Figure 1. Schematic and band structure of an encapsulated InSe device.** (a) Optical microscope picture of the 3L device with layers highlighted in false color. Few-layer graphene (FLG), InSe and hBN layers are depicted in grey, violet and blue, respectively. (b) Three-dimensional schematic representing an hBN-encapsulated InSe layer with graphene contacts and a wide back gate. The sample is optically excited using a red laser and excitonic species are formed within the semiconductor. When a gate voltage is applied, holes in InSe can tunnel through the bottom hBN giving rise to a tunneling photocurrent. The electrical measurement scheme is shown on the right. (c) Band alignment of the materials of interest in the device of Figure 1b. Two electron donor (red) and one acceptor (blue) states are induced by the presence of selenium vacancies. (d) Band structure of 3-layer slab of InSe with a selenium vacancy A van Hove singularity arises due to the band flattening at the $\Gamma$ point. (e) The density of states in a 3L-InSe shows a sharp peak at the valence band maximum, as well as the donor (red) and acceptor (blue) states. The inset shows in logarithmic scale the region between the highest donor state and the beginning of the conduction band, with its minimum located at 1.54 eV with respect to the acceptor state.

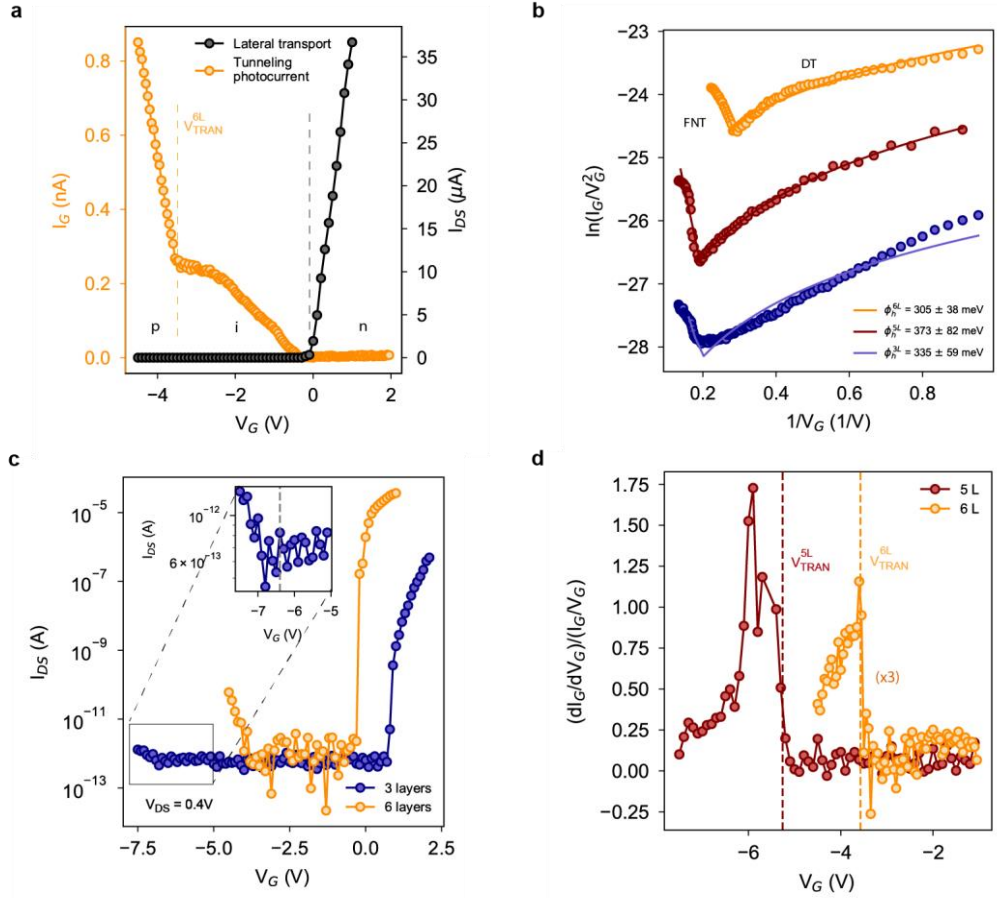

**Figure 2. Field-effect transport and tunneling photocurrent with few-layer InSe.** (a) The gate-dependent lateral current between the FLG contacts in a 6L device (black) shows predominantly n-type conduction in linear scale. The dashed grey line indicates the onset of n-type conduction. A gate-dependent tunneling photocurrent is induced when the sample is illuminated using a 633 nm laser at 50 μW of power (yellow). The behavior of the tunneling current can be divided into two main regions, which are separated by a yellow dashed line with respect to the gate voltage (b) The gate-dependent tunneling photocurrent data for 3L, 5L and 6L devices are reported blue, red and yellow dots, respectively, using a $ln(I_G/V_G^2) - 1/V_G$ scale. The tunneling behavior is modeled using Simmons' approximation, with direct (DT) and Fowler-Nordheim tunneling (FNT) regimes separated by a sharp onset. FNT fittings reveal effective barriers for tunneling holes of 237 meV, 340 meV and 292 meV for 3L, 5L and 6L samples, respectively. (c) Field-effect lateral transport in the 3L (measured at 80mK) and 6L devices in logarithmic scale. Sub-threshold p-type conduction was achieved in both samples. Note a lower subthreshold slope in the 3L with respect to the 6L device (Supplementary Note 2). The inset highlights the onset of p-type conduction in the 3L device ($V_{ON}^{3L} = -6.4V$. (d) Ratio between the differential conductance and the tunneling conductance at low temperature for 5L and 6L devices. The weaker 3L signal does not allow us to observe changes in the normalized conductance at cryogenic temperature. Further discussion on the temperature dependence of the $dI_G/dV_G$ signals is provided in Supplementary Note 3.

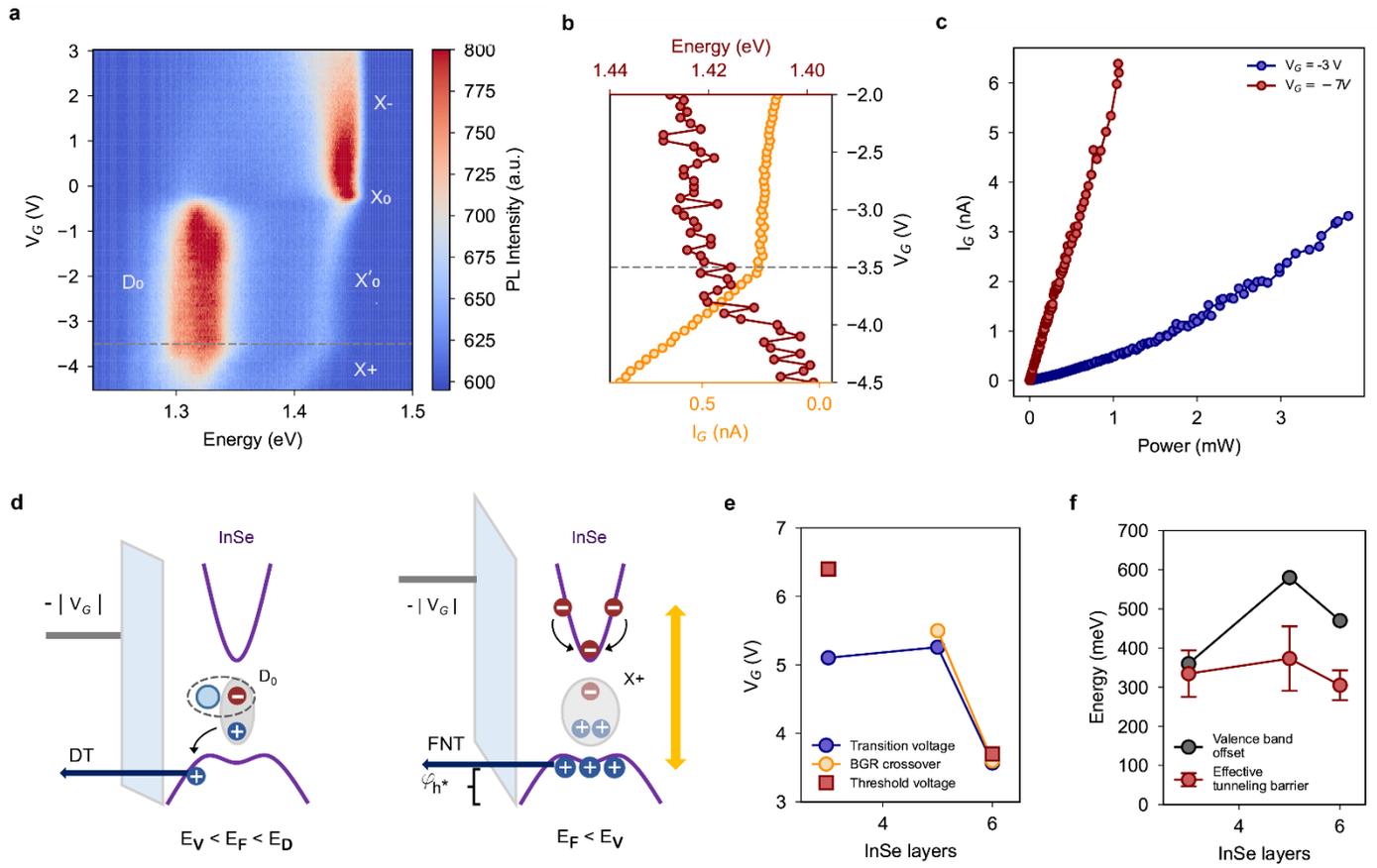

**Figure 3. Excitonic emission and tunneling mechanisms with few-layer InSe.** (a) Gate-dependent PL spectra for the 6L device. The change in the main excitonic peaks with respect to gate voltage is related to different charge configurations (Supplementary Note 4). When the Fermi level in the semiconductor enters the valence band (grey dashed line), the highest-energy peak shifts linearly with respect to the gate voltage due to bandgap renormalization in the presence of a dense hole Fermi sea[20]. (b) Peak position of the highest-energy excitonic species (red) and tunneling photocurrent (yellow) in the vicinity of the VBM. The DT-to-FNT transition ($V_{TRAN}^{5L}$) and the BGR start ($V_{BGR}^{5L}$) are highlighted by dashed yellow and red lines, respectively. (c) Power-dependent tunneling photocurrent for the 5L device (Supplementary Note 5) when the Fermi level lies within the bandgap (blue) or has entered the valence band (red). The superlinear behavior of the former case indicates that second-order excitonic effects play a role, which is excluded within the valence band. (d) Auger and exciton-exciton recombination mechanisms can radiatively transfer their energy to resident carriers and induce hot-hole tunneling through a thick hBN barrier in the presence of a vertical electric field ($E_V < E_F < E_D$). When the Fermi level lies within the valence band ($E_F < E_V$), excitonic species are subject to interactions with the Fermi sea induced by the high DOS, and directly photoexcited holes are responsible for the photocurrent by FNT through the hBN barrier (right). (e) Transition voltage ($V_{TRAN}$), BGR crossover ($V_{BGR}$) and p-type onset ($V_{ON}$) for the 3L, 5L and 6L devices. The 5L sample was not equipped with two FLG contacts for lateral transport measurements, and the intensity of PL emission in the 3L sample was too low to be analyzed as a function of gate voltage. (f) Layer-dependent effective tunneling barrier extracted from FNT fittings (Figure 2b) and VBOs obtained from DFT calculations (Supplementary Figure S1).

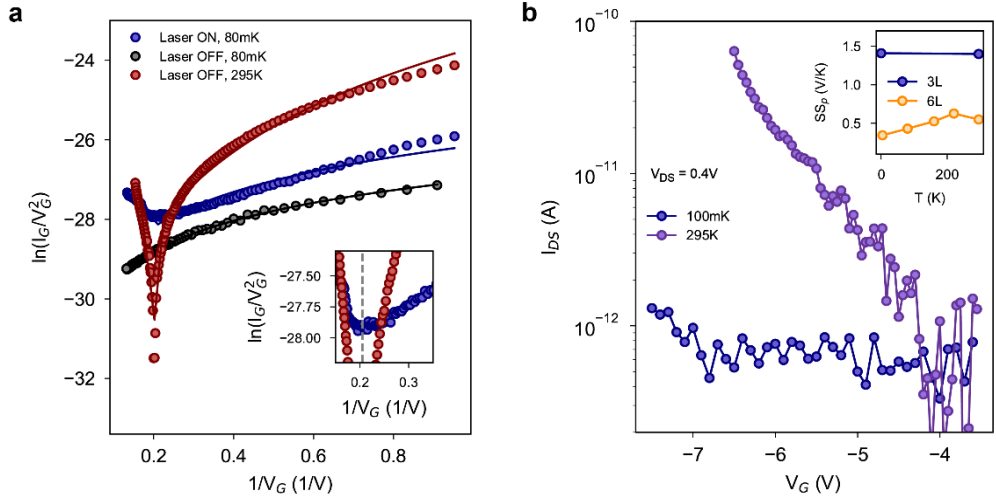

**Figure 4. Temperature-dependent lateral transport and tunneling current.** (a) Tunneling current in the 3L device without laser illumination at 100 mK (black) and 295K (red). The logarithmic dependence in the $ln(I_G/V_G^2) - 1/V_G$ reveals no transition in the dark at low temperatures and a sharp transition at room temperature. In the inset, the tunneling photocurrent at 100 mK (blue) is compared with the dark tunneling current at room temperature, revealing a comparable trend in the Fowler-Nordheim regime ($1/V_G < 0.2V$ and a transition voltage of around 5 V for both cases. (b) Lateral p-type transport for the 3L device at 100 mK (blue) and 295 K (violet). The thermal broadening due to Fermi-Dirac scaling with temperature induces a shift in the first-detectable subthreshold p-type signal[44], as further discussed in Supplementary Note 2. However, no significant change in the subthreshold slope is detected. In the inset, the p-type subthreshold slopes ($SS_p$) of the 3L (blue) and 6L (yellow) devices are shown. The significant difference in magnitude of $SS_p^{3L}$ and $SS_p^{6L}$ both at room and cryogenic temperatures is related to the increase in hole effective mass in InSe for decreasing number of layers.